\documentstyle[12pt]{article}

0

\begin{document}
\begin{flushright}
hep-th/9412177
\end{flushright}
\vglue 1cm

\begin{center}

{}~\vfill

{\large \bf  THE ONE-LOOP DIVERGENCES OF \protect \\
{}~\vfill
THE LINEAR GRAVITY WITH THE TORSION TERMS}

\vfill

{\large M. Yu. Kalmykov}
\footnote { E-mail: $kalmykov@thsun1.jinr.dubna.su$  \protect \\
Supported in part by RFFR grant \# 94-02 03665-a and ISF grant \#
RFL300}

\vspace{1cm}

{\em Bogoliubov Laboratory of Theoretical Physics,
 Joint Institute for Nuclear  Research,
 $141~980$ Dubna $($Moscow Region$)$, Russian Federation} \protect \\

\vspace{1cm}

and

\vspace{1cm}

{\large P. I. Pronin}\footnote{E-mail: $pronin@theor.phys.msu.su$}

\vspace{1cm}

{\em Department of Theoretical Physics, Physics Faculty\\
 Moscow State University, $117234$, Moscow, Russian  Federation}

\end{center}

\vfill

\begin{abstract}

We investigate the role of the torsion field at the quantum level.
One-loop counterterms are calculated in the theory with terms quadratic
in the torsion field. We have shown that the theory is finite at the
one-loop level.

\end{abstract}

\vfill

\pagebreak

\section{Introduction}

The construction of a quantum theory of gravity is an unresolved problem
of modern theoretical physics. It is well know that the Einstein theory
of gravity is not renormalizable  in an ordinary sense
\cite{tHV,GS,vandeven}.  Therefore, one  needs to modify
the theory or to show that the difficulties presently encountered in
the theory are only artifacts of perturbation theory. The simplest
method of modifying the Einstein theory is to introduce terms quadratic
in the curvature tensor in the action of the theory.
This theory is renormalizable and asymptotically  free
 but it is not unitary because the ghosts and tachyons are present in
the spectrum of the theory ~\cite{Stelle,Avramidi}.
It should be noted that the unitarity of the theory cannot be restored
by means of loop corrections or adding an interaction with matter fields
{}~\cite{AT,J}.  Hence, one needs to use a new method in order to
construct a theory of gravity.

Among  various methods of constructing a quantum theory of
gravity one should emphasize the gauge approach as the most
promising ~\cite{H-76,DD,M,H-85}.
In the gauge treatment of gravity there are two sets of dynamical
variables, namely, the vierbein $h^a_{~\mu} (x)$ and local Lorentz
connection $\omega^a_{~b\mu }(x)$ or metric $g_{\mu  \nu }(x)$ and
affine connection $\Gamma^\sigma _{~\mu \nu }(x)$. The theory based on
the first set of variables is called the Poincar\'{e} gauge
gravitational theory with the structure group $P_{10}$
{}~\cite{T}. A curvature tensor $R^a_{~b \mu  \nu }(\omega )$ and
a torsion tensor $Q^a_{~\mu  \nu }(h,\omega)$, which are the strength
tensors of the Poincar\'{e} gauge gravitational theory, are defined by
the following relations:

\begin{eqnarray}
R^a_{~b \mu \nu }(\omega) & =  & \partial_\mu \omega^a_{~b \nu
} - \partial_\nu \omega^a_{b \mu }  + \omega^a_{~c \mu } \omega^c_{b
\nu } - \omega^a_{~c \nu } \omega^c_{b \mu }
\nonumber \\
Q^a_{~\mu  \nu }(h,\omega ) & = & -\frac{1}{2}
\left(\partial_\mu h^a_{~\nu } -
\partial_\nu h^a_{~\mu } + \omega^a_{~c \nu } h^c_{~\mu } -
\omega^a_{~c \mu } h^c_{~\nu } \right)
\nonumber
\end{eqnarray}

The theory based on the second set of variables is called the affine
gauge gravitational theory with the structure gauge group
$GA(4,R)$ ~\cite{Lord,Neem-88}.
The strength tensor of the theory is the curvature tensor
$\tilde R^\sigma _{~\lambda  \mu \nu }(\Gamma )$ defined as:

\begin{equation}
\tilde R^\sigma _{~\lambda  \mu  \nu }(\Gamma ) = \partial_\mu
\Gamma^\sigma _{~\lambda \nu }  - \partial_\nu \Gamma^\sigma _{\lambda
\mu } + \Gamma^\sigma_{~\alpha \mu } \Gamma^\alpha_{~\lambda  \nu } -
\Gamma^\sigma_{~\alpha  \nu }  \Gamma^\alpha_{\lambda \mu }
\end{equation}
The Lagrangian of a gauge theory is  built out of
terms quadratic in the strength tensor of fields. In the Poincar\'{e}
or affine gauge theories the Lagrangians are defined as

\begin{eqnarray}
{\it L}_{P_{10}} & = &
\left(\frac{A_i}{{\it k}^2}Q^2(h,\omega ) + B_j R^2(\omega) \right) h
\nonumber \\
{\it L}_{GA(4,R)} & = &
C_j \tilde R^2(\Gamma ) \sqrt{-g}
\label{GA}
\end{eqnarray}
where $A_i, B_j$ and $C_j$ are arbitrary constants, and $R^2$ and $Q^2$
are now a symbolic notation for the contractions of the curvature or the
torsion tensors, respectively.

For the classical limit, coinciding with the Einstein theory, to exist
one needs to add  a term linear in the curvature tensor to the Lagrangian.

At the present time, there are a lot of papers concerning the classical
problems of the Poincar\'{e} and affine gauge gravitational
theories ~\cite{HS,HKH-76,HLN-77,H-89,M-93}.
However, the renormalizability properties of the theories have been
insufficiently studied ~\cite{Martel,D,shapiro,LN,MKL}.

In the above-mentioned theories, the torsion tensor has a different
meaning. In the Poincar\'{e} gauge gravitational theory, the torsion
tensor is a strength tensor of the tetrad fields. Hence, the terms
quadratic in the torsion field  must be present in the Lagrangian of
the theory. In the affine gauge theory
of gravity, the torsion tensor plays an auxiliary role. Therefore, the
Lagrangian of the affine-metric theory of gravity  can not contain
terms quadratic in the torsion field.

The main goal of our work is to investigate the role of the torsion
field in the affine-metric theory of gravity at the quantum level.
The consideration of the full affine gauge theory described by the
action (\ref{GA}) is very cumbersome and technically complex.
To understand the role of the torsion fields in the affine-metric
theory we will consider a simple model with the terms quadratic in the
torsion fields. The Lagrangian of the model is the following:

\begin{equation}
{\it S}_{gr}  =  - \frac{1}{{\it k}^2} \int d^4 x \sqrt{-g}
\Bigl( \tilde R(\Gamma ) - 2 \Lambda
 + b_1 Q_{\sigma \mu \nu} Q^{\sigma \mu \nu} + b_2
Q_{\sigma \mu \nu} Q^{\nu \sigma \mu} + b_3 Q_\sigma Q^\sigma
\Bigr)
\label{action}
\end{equation}
where
$\Lambda $ is a cosmological constant and $\{b_i\}$ are
arbitrary constants.

We consider $\Gamma^\sigma _{~\mu \nu }(x)$ and $g_{\mu \nu }(x)$
as independent dynamical fields.

This model is not an affine gauge theory in the above-mentioned sense.
However, some properties of the affine metric theory can be studied by
means of the model (\ref{action}). In particular, this model possesses
the same symmetries as the affine gauge theory.

The main obstacle to the renormalizability of Einstein's theory
consists in the existence of the dimensional constant (Newton's
constant) and thus in the need for new counterterms in each order of
perturbation theory. In other words, renormalizable quantum
gravity in four-dimensional space-time must contain terms with
dimension four. However, the presence of an additional symmetry in the
theory may improve the renormalization properties of the theory. For
example, because of the presence of supersymmetry the terms violating
the renormalizability of supergravity show up only in higher loops. The
considered model (\ref{action}) has the Lagrangian of dimension two.
Hence, this theory is not renormalizable in all orders of perturbation
theory.  However, the projective invariance ~\cite{Sand,HKMM-91}
existing in the model with metric and affine connection as independent
dynamical variables may influence the renormalizability of the theory
{}~\cite{MKL2}.

In the present work we will research the following problems:

\begin{enumerate}
\item The role of the terms quadratic in the torsion fields at the
quantum level in the theory with independent metric and affine fields.
\item The influence of an additional projective symmetry on the
one-loop counterterms.
\item The influence of the terms quadratic in the torsion fields on the
one-loop renormalizability of the theory.
\end{enumerate}

We use the following notation:
$$ c = \hbar = 1;~~~~~ \mu , \nu  = 0,1,2,3;~~~~~ {\it k}^2 = 16 \pi G
{}~~~~~\varepsilon = \frac{4 - n}{2}$$
$$ \tilde R_{\mu \nu }(\Gamma ) =
\tilde R^\sigma_{~\mu \sigma \nu }(\Gamma ),~~~~~
 \tilde R(\Gamma ) = \tilde R_{~\mu \nu }(\Gamma ) g^{\mu \nu },
{}~~~~~g = det( g_{\mu \nu }) $$

The objects marked by the tilde $\tilde{} $  are constructed by means of
the affine connection $\Gamma^\sigma_{~\mu \nu }$. The others are the
Riemannian objects.

\section{Symmetries of the model and equations of motion}

Let us consider the classical symmetries of the model (\ref{action}).
This model is invariant under the general coordinate transformation

\begin{eqnarray}
x^\mu & \rightarrow &  'x^\mu  = x^\mu + {\it k} \xi^\mu (x)
\nonumber \\
g_{\mu  \nu }(x) & \rightarrow & 'g_{\mu \nu }(x)  = g_{\mu \nu }(x)
- {\it k} \partial_\mu \xi^\alpha  g_{\alpha \nu }(x)
- {\it k} \partial_\nu \xi^\alpha  g_{\alpha \mu }(x)
- {\it k} \xi^\alpha \partial_\alpha  g_{\mu \nu }(x) + O({\it k}^2)
\nonumber \\
\Gamma^\sigma _{~\mu \nu }(x) & \rightarrow  &
'\Gamma^\sigma _{~\mu \nu }(x)  = \Gamma^\sigma _{~\mu \nu }(x)
- {\it k} \partial_\mu \xi^\alpha  \Gamma^\sigma_{~\alpha \nu }(x)
- {\it k} \partial_\nu \xi^\alpha  \Gamma^\sigma_{~\mu \alpha }(x)
\nonumber \\
&& ~~~~~~~~~~~
+ {\it k} \partial_\alpha   \xi^\sigma \Gamma^\alpha_{~\mu \nu }(x)
- {\it k} \xi^\alpha \partial_\alpha \Gamma^\sigma_{~\mu \nu }(x)
- {\it k} \partial_{\mu \nu } \xi^\sigma + O ({\it k}^2)
\label{coordinate}
\end{eqnarray}

Moreover, in the case of special choice of the coefficients $\{b_j\}$,
the action (\ref{action}) is invariant under the following
transformation of fields:

\begin{eqnarray}
x^\mu & \rightarrow &  'x^\mu  = x^\mu \nonumber \\
g_{\mu  \nu }(x) & \rightarrow & 'g_{\mu \nu }(x)  = g_{\mu \nu }(x)
\nonumber \\
\Phi_{mat}(x) & \rightarrow & '\Phi_{mat}(x)  =  \Phi_{mat}(x)
\nonumber \\
\Gamma^\sigma _{~\mu \nu }(x) & \rightarrow  &
'\Gamma^\sigma _{~\mu \nu }(x)  = \Gamma^\sigma _{~\mu \nu }(x)
+ {\it k} \delta^\sigma_\mu C_\nu(x)
\label{projective}
\end{eqnarray}
where $C_\nu (x)$ is an arbitrary vector.

This is the projective transformation ~\cite{Sand},~\cite{HKMM-91}.

It is easy to show that under the projective transformations
(\ref{projective}) the curvature and torsion tensors transform in
the following way:

\begin{eqnarray}
\tilde R^\sigma_{~\lambda \mu \nu}(\Gamma ) & \rightarrow &
'\tilde R^\sigma_{~\lambda \mu \nu}('\Gamma ) =
\tilde R^\sigma_{~\lambda \mu \nu}(\Gamma)  + \delta^\sigma_\lambda
{\it k} \left(\partial _\mu C_\nu - \partial_\nu C_\mu \right)
\nonumber \\
Q^\sigma_{\mu \nu } & \rightarrow & 'Q^\sigma_{\mu \nu }  =
Q^\sigma_{\mu \nu }  + \frac{1}{2} {\it k} \left( \delta^\sigma_\mu
C_\nu - \delta^\sigma_\nu C_\mu \right)
\nonumber
\end{eqnarray}

Hence, the action (\ref{action}) is invariant under
the projective transformation (\ref{projective}) at the tree level
only under the condition

\begin{equation}
2b_1 + 3b_3 - b_2 = 0
\label{cond}
\end{equation}

The classical fields $ç^\sigma_{~\mu \nu}$ and $g_{\mu \nu}$ satisfy the
following equations of motion:

\begin{eqnarray}
\frac{\delta {\it S}_{gr}}{\delta g^{\mu \nu}}
& = &
\tilde R_{(\mu \nu)} - \frac{1}{2} \tilde R g_{\mu \nu} + \Lambda g_{\mu \nu}
\nonumber \\
& - & b_1 Q_{\mu \alpha \beta} Q_\nu^{~\alpha \beta} +
2 b_1 Q_{\alpha \beta \mu} Q^{\alpha \beta}_{~~\nu}
 - b_2 Q^{\alpha b}_{~~\mu}Q_{\beta \alpha \nu} +
b_3 Q_\mu Q_\nu
\nonumber \\
& + & \frac{1}{2}~g_{\mu \nu}
\left( b_1~ Q_{\sigma \alpha \beta} Q^{\sigma \alpha \beta}
+ b_2~Q_{\sigma \alpha \beta} Q^{\beta \sigma \alpha}
+ b_3~Q_\sigma Q^\sigma  \right ) = 0
\label{mass-shell1}
\end{eqnarray}

and

\begin{eqnarray}
\frac{\delta {\it S}_{gr}}{\delta \Gamma^\sigma_{~\alpha \beta}} &=&
D^{\alpha \lambda}_{~~\lambda} \delta^\beta_\sigma +
D_\sigma g^{\alpha \beta}  -
\biggl(1 - \frac{1}{2} b_2\biggr) D^{\alpha \beta}_{~~\sigma} -
\biggl(1 - \frac{1}{2}b_2\biggr) D^{\beta ~ \alpha}_{~ \sigma}
\nonumber \\
& + & b_1 D_\sigma^{~\alpha \beta} - b_1 D_\sigma^{~\beta \alpha} -
\frac{1}{2}b_2 D^{\alpha ~ \beta}_{~ \sigma} -
\frac{1}{2}b_2 D^{\beta \alpha}_{~~\sigma} \nonumber \\
& + & \frac{1}{2} b_3 D^\alpha \delta^\beta_\sigma  -
\frac{1}{2} b_3  D^\beta \delta^\alpha_\sigma -
\frac{1}{2}b_3 D_\lambda^{~\lambda \alpha} \delta^\beta_\sigma
+ \frac{1}{2}b_3 D_\lambda^{~\lambda \beta} \delta^\alpha_\sigma = 0
\label{mass-shell2}
\end{eqnarray}
where
\begin{eqnarray}
D^\sigma_{~\mu \nu } & = & \Gamma^\sigma_{~\mu \nu } -
g^{\sigma \lambda } \frac{1}{2} \left(- \partial_\lambda g_{\mu \nu } +
\partial_\mu  g_{\nu \lambda } + \partial_\nu  g_{\mu \lambda }\right)
\nonumber \\
D_\sigma & = & D^\alpha_{~\sigma \alpha}
\nonumber
\end{eqnarray}

Equation  (~\ref{mass-shell2}) has two solutions

\begin{enumerate}
\item if  $2b_1 + 3b_3  - b_2 = 0$,
\begin{equation}
D^\sigma_{~\mu \nu} = \delta^\sigma_\mu C_\nu(x)
\label{defect1}
\end{equation}
where $C_\nu $ is an arbitrary vector.

\item  if $2b_1 + 3b_3  - b_2 \neq 0$
\begin{equation}
D^\sigma_{~\mu \nu} = 0
\label{defect2}
\end{equation}

\end{enumerate}

Taking into account (\ref{defect1}) or (\ref{defect2}) we
obtain from (\ref{mass-shell1})

\begin{equation}
R_{\mu \nu} = \Lambda g_{\mu \nu}
\end{equation}

In the next chapter, we will consider at the quantum level two cases:
\begin{itemize}
\item the theory without the projective invariance (the condition
(\ref{cond}) is not satisfied.)
\item the theory with the projective invariance (the condition
(\ref{cond}) is fulfilled)
\end{itemize}

\section{One-loop counterterms}

For calculating the one-loop effective action we will use the background
field method ~\cite{BDW-67,BDW} and the Schwinger-DeWitt
technique ~\cite{G,bar}. In the gauge theories, the
renormalization procedure may violate the gauge invariance at the
quantum level, thus destroying the renormalizability of the theory.
Therefore, one is bound to apply an invariant renormalization. We will
use the dimensional regularization and minimal subtraction scheme in our
loop calculation.  This is the invariant renormalization.

In accordance with the background field
method, all dynamical variables are rewritten as a sum of
classical and quantum parts. In general case, the dynamical variables
in the affine-metric theory are
$\underline{\Gamma}^\sigma_{\mu \nu }, \bar g_{\mu
\nu } = g_{\mu \nu }(-g)^r$ or $\bar g^{\mu \nu } = g^{\mu \nu }
(-g)^s$, where $r,s$ are the numbers satisfying the only condition:  $r
\neq -\frac{1}{4}, s \neq \frac{1}{4}$. The one-loop counterterms on
the mass-shell do not depend on the value of $r$ and $s$.  To simplify
our calculation, we use the following numbers $r = s = 0$.

The fields $\underline{\Gamma}^\sigma_{~\mu \nu }$ and
$\underline{g}_{\mu \nu } $ are now rewritten according to

\begin{eqnarray}
\underline{\Gamma}^\sigma_{~\mu \nu } & = & \Gamma^\sigma_{\mu \nu }
+ {\it k} \gamma^\sigma_{~\mu \nu }  \nonumber \\
\underline{g}_{\mu \nu }  & = & g_{\mu \nu } + {\it k}h_{\mu \nu }
\label{expansion}
\end{eqnarray}
where $\Gamma^\sigma_{~\mu \nu } ,g_{\mu \nu }$
are the classical parts satisfying equations (\ref{mass-shell1}) and
(\ref{mass-shell2}).

The action (\ref{action}) expanded as a power series in the quantum
fields (\ref{expansion}) defines the effective action for calculating
the loop counterterms. The one-loop effective Lagrangian quadratic
in the quantum fields is

\begin{equation}
{\it L}_{eff} = - \frac{1}{2} \gamma^ \sigma_{~\mu \nu}
\tilde F_{~\sigma~~\lambda}^{~\mu \nu~\alpha \beta}
\gamma^\lambda_{~\alpha \beta}
  - \frac{1}{2} h^{\alpha \beta} ~X_{\alpha \beta \mu \nu} ~h^{\mu \nu}
-  \gamma^\lambda_{~\alpha \beta}  B_{\lambda ~~~\mu
\nu}^{~\alpha \beta \sigma} \nabla_\sigma h^{\mu \nu}
+ \gamma^\sigma_{~\mu \nu} ~H^{~\mu \nu}_{\sigma ~~\alpha \beta}
{}~h^{\alpha \beta}
\label{expan}
\end{equation}
where

\begin{eqnarray}
\tilde F_{\alpha ~~\mu}^{~\beta \lambda ~ \nu \sigma} & = &
g^{\beta \lambda} \delta^\nu_\alpha \delta^\sigma_\mu +
g^{\nu \sigma}\delta^\lambda_\alpha \delta^\beta_\mu -
\biggl(1 - \frac{b_2}{2} \biggr)
g^{\beta \sigma} \delta^\nu_\alpha \delta^\lambda_\mu -
\biggl(1 - \frac{b_2}{2} \biggr)
g^{\lambda \nu}\delta^\sigma_\alpha \delta^\beta_\mu
\nonumber \\
& + &  b_1~ g_{\alpha \mu} g^{\beta \nu} g^{\sigma \lambda} -
b_1~g_{\alpha \mu} g^{\lambda \nu} g^{\beta \sigma}   -
\frac{b_2}{2}g^{\lambda \sigma} \delta^\beta_\mu \delta^\nu_\alpha
 - \frac{b_2}{2} g^{\beta \nu}\delta^\lambda_\mu \delta^\sigma_\alpha
 \nonumber \\
& + &  \frac{b_3}{2}g^{\beta \nu}\delta^\lambda_\alpha
\delta^\sigma_\mu - \frac{b_3}{2}g^{\beta \sigma
 }\delta^\nu_\mu\delta^\lambda_\alpha - \frac{b_3}{2}g^{\lambda
 \nu}\delta^\beta_\alpha\delta^\sigma_\mu + \frac{b_3}{2}g^{\sigma
 \lambda}\delta^\beta_\alpha \delta^\nu_\mu \end{eqnarray}

\begin{eqnarray}
P^{ \alpha \beta \mu \nu } & = & \frac{1}{4} \left(
g^{\alpha \mu } g^{\beta \nu } + g^{\alpha \nu }g^{\beta \mu } -
g^{\alpha \beta }g^{\mu \nu } \right)
\nonumber \\
B_{\lambda~~~(\mu \nu)}^{~\alpha \beta \sigma} & = &
2 \left(
\delta^\sigma_\lambda P^{\alpha \beta}_{~~ \mu \nu}
- \delta^\beta_\lambda P^{\alpha \sigma}_{~~\mu \nu} \right)
\nonumber \\
X_{((\alpha \beta) ( \mu \nu))} & = & 2R_{\alpha \mu} g_{\beta \nu} -
\left(R - 2 \Lambda \right) P_{\alpha \beta \mu \nu} -
R_{\alpha \beta} g_{\mu \nu}
- \frac{1}{2} b_1 ~g_{\alpha \beta} ~Q_{\mu \sigma \lambda} ~Q_{\nu}^{~
\sigma \lambda}
\nonumber \\
& + & b_1 \left(
2 Q^\sigma_{~ \alpha \mu}  ~Q_{\sigma \beta \nu}
- 4 Q_{\alpha \mu \sigma} ~Q_{\beta \nu}^{~~\sigma}
+ 4 Q^{\sigma \lambda}_{~~\alpha} ~Q_{\sigma \lambda \mu} ~g_{\beta
\nu} - g_{\alpha \beta} ~Q^{\sigma \lambda}_{~~\mu} ~Q_{\sigma \lambda
\nu} \right)
\nonumber \\
& + & b_2 \left(
 2 g_{\beta \mu} ~Q^{\sigma \lambda}_{~~\alpha}
{}~Q_{\lambda \nu \sigma}
- g_{\alpha \beta} ~Q^{\sigma \lambda}_{~~\mu} ~Q_{\lambda
\nu \sigma} \right)
+ b_3 \left(
g_{\alpha \beta} ~Q_\mu ~Q_\nu - 2 g_{\alpha \mu} ~Q_\beta ~Q_\nu
\right)
\nonumber \\
& - & P_{\alpha \beta \mu \nu } \left(
b_1 Q^{\lambda \rho \tau} Q_{\lambda \rho \tau} + b_2 Q_{\lambda
\rho \tau} Q^{\tau \lambda \rho} + b_3 Q^\lambda Q_\lambda \right)
\nonumber \\
H_{\sigma ~~(\alpha \beta)}^{~ \mu \nu}
& = &
2 \left(
D^\mu_{~\rho \tau} \delta^\nu_\sigma + D_\sigma \delta^\mu_\rho
\delta^\nu_\tau - D^\mu_{~\rho \sigma} \delta^\nu_\tau -
D^\nu_{~\sigma \tau} \delta^\mu_\rho \right)
P^{\rho \tau}_{~~ \alpha \beta}
\nonumber \\
& + & 2 b_1 \left(
Q_{\sigma~\alpha}^{~\mu} ~\delta^\nu_\beta
- Q_{\sigma~\alpha}^{~ \nu} ~\delta^\mu_\beta
- Q_\alpha^{~\mu \nu} g_{\beta \sigma}
- \frac{1}{2} Q_\sigma^{~\mu \nu} g_{\alpha \beta} \right)
\nonumber \\
& + & 2 b_2 \left(
Q^\nu_{~ \sigma \lambda} P^{\mu \lambda}_{~~ \alpha \beta}
- Q^\mu_{~ \sigma \lambda} P^{\nu \lambda}_{~~ \alpha \beta} \right)
 +  2 b_3 \left(
\delta^\nu_\sigma Q_\lambda P^{\mu \lambda}_{~~ \alpha \beta}
- \delta^\mu_\sigma Q_\lambda P^{\nu \lambda}_{~~ \alpha \beta}
\right)
\nonumber
\end{eqnarray}

Parentheses around index pairs denote symmetrization while parentheses
around four indices mean symmetrization also under pair interchange.
These symmetries are automatically enforced by the symmetries of the
quantum fields multiplying these quantities.

Let us consider the first case: the theory without the
projective invariance (the condition (\ref{cond}) is not
satisfied).  To get the diagonal form of the effective Lagrangian we
are to replace the dynamical variables in the following way:

\begin{equation}
\tilde \gamma^\sigma_{~\mu \nu}  =  \gamma^\sigma_{~\mu \nu} +
 F^{-1 \sigma~~ \lambda}_{~~~~ \mu \nu~ \alpha \beta}
 \left(
 B^{~ \alpha \beta \tau}_{\lambda ~~~\rho \epsilon}
 \nabla _ \tau  -  H_{\lambda~~\rho \epsilon }^{~\alpha \beta}
 \right) h^{\rho  \epsilon}
\label{replace}
\end{equation}
where $F^{-1 \sigma~~ \lambda}_{~~~~ \mu \nu~ \alpha \beta}$ is
the propagator of the quantum field $\gamma^\sigma_{~\mu \nu }$
satisfying two conditions

\begin{equation}
F^{-1 \sigma ~~\lambda}_{~~~~\mu \nu ~\alpha \beta} =
F^{-1 \lambda ~~\sigma}_{~~~~\alpha \beta ~\mu \nu}
\label{first}
\end{equation}

\begin{equation}
F^{-1 \sigma ~~\lambda}_{~~~~\mu \nu ~\alpha \beta}
\tilde F_{\lambda~~\rho}^{~\alpha \beta ~ \tau \epsilon} =
\delta^\sigma_\rho \delta^\tau_\mu \delta^\epsilon_\nu
\label{second}
\end{equation}

Having solved equations (\ref{first}) and (\ref{second}) we obtain the
following result:

\begin{eqnarray}
F^{-1 \alpha ~~\mu}_{~~~~\beta \sigma ~\nu\lambda}  =
 & - & \frac{1}{4}  g^{\alpha \mu} g_{\beta \sigma} g_{\nu \lambda} +
  A_2 g^{\alpha \mu} g_{\beta \nu} g_{\sigma \lambda}
+ \biggl( \frac{1}{12} - \frac{1}{3} \left(A_2 - A_1 \right)
\biggr)
g_{\nu \beta} \delta^\mu_\lambda \delta^\alpha_\sigma
\nonumber \\
& + & \frac{1}{4}  \biggl( g_{\nu \lambda} \delta^\mu_\beta
\delta^\alpha_\sigma
+ g_{\beta \sigma} \delta^\alpha_\nu \delta^\mu_\lambda \biggr)
 + A_1 \biggl( g_{\nu \sigma}
\delta^\alpha_\lambda \delta^\mu_\beta + g_{\beta \lambda}
\delta^\mu_\sigma \delta^\alpha_\nu \biggr)
\nonumber \\
& + &
\frac{1}{4}
\biggl( g_{\nu \lambda}
\delta^\mu_\sigma \delta^\alpha_\beta + g_{\beta \sigma}
\delta^\alpha_\lambda \delta^\mu_\nu \biggr)
- \left(A_2 - \frac{1}{2} \right)
g^{\alpha \mu } g_{\sigma \nu } g_{\beta \lambda }
\nonumber \\
& + &
\biggr( \frac{1}{3} \left(A_2 - A_1 \right) - \frac{1}{4} \biggr)
\biggl( g_{\beta \lambda}
\delta^\mu_\nu \delta^\alpha_\sigma + g_{\beta \nu} \delta^\alpha_\beta
\delta^\mu_\lambda \biggr) \nonumber \\
\nonumber \\
& + &
\biggl( \frac{1}{12}  - \frac{1}{3} \left(A_2 - A_1 \right) +
\frac{2}{3(2b_1 + 3b_3 - b_2)} \biggr)
g_{\sigma \lambda} \delta^\mu_\nu \delta^\alpha_\beta
\nonumber \\
& - &
\left( \frac{1}{2}  + A_1 \right)
g_{\nu \beta } \delta^\mu_\sigma \delta^\alpha_\lambda  -
\left( \frac{1}{2}  + A_1 \right)
g_{\sigma \lambda } \delta^\mu_\beta \delta^\alpha_\nu
\end{eqnarray}
where the constants $A_2$ and $A_1$ are defined by the following
expressions :

\begin{eqnarray}
A_2 & = & \frac{2b_2^2 - 4b_1^2 - 2b_1 b_2 - 6b_2 - 2b_1 + 4}{d}
\nonumber \\
A_1 & = & - A_2 - \frac{4b_1}{d} + \frac{8}{d}
\label{coef}
\end{eqnarray}

where

\begin{equation}
d \equiv 8(b_2^2 - 2 b_2 + 1 + b_1 - b_1 b_2 - 2 b_1^2) \neq 0
\label{d}
\end{equation}

The replacement (\ref{replace}) does not change the functional measure

$$ {\it det} \left|
\frac{\partial (h, ~\tilde \gamma )}{\partial (h,~
\gamma)}  \right| = 1 $$

We violate the coordinate invariance of the action (\ref{expan}) by
means of the following gauge:

\begin{equation}
F_\mu = \nabla_\nu h^\nu_{~\mu} - \frac{1}{2} \nabla_\mu
h^\alpha_{~\alpha}
\label{firstghost}
\end{equation}

\begin{equation}
{\it L}_{gh} = \frac{1}{2} F_\mu F^\mu
\label{secondghost}
\end{equation}

The Lagrangian of the coordinate ghost is

\begin{equation}
{\it L}_{gh} = - \overline c^\mu \biggl(
g_{\mu \nu } \nabla^2 + R_{\mu \nu } \biggr) c^\nu
\label{coorghost}
\end{equation}

We don't give the details of cumbersome calculations.
The one-loop counterterms on the mass-shell including
the contributions of the quantum and ghost fields are

\begin{equation}
\triangle \Gamma^1_\infty = - \frac{1}{32 \pi^2 \varepsilon}
\int d^4x \sqrt{-g} \Biggl(
\frac{53}{45}
R_{\alpha \beta \mu \nu}R^{\alpha \beta \mu \nu}
 - \frac{58}{5} \Lambda^2 \Biggr)
\label{result}
\end{equation}

Let us consider the second case: the theory possessing the projective
invariance (the condition (\ref{cond}) is fulfilled).
In this case, the propagator
$ F^{-1 \alpha ~~\mu}_{~~~~\beta \sigma ~\nu\lambda} $
of the quantum field
$\gamma^\sigma_{\mu \nu}$  does not exist because of the projective
invariance of the effective Lagrangian (\ref{expan}).

We consider the projective invariance as a gauge symmetry. Hence, we
must fix this symmetry at the quantum level. The gauge fixing Lagrangian
is

\begin{equation}
{\it L}_{gf}  = b^\mu F_\mu + \pi^\mu f_\mu - \frac{1}{2} b^\mu b_\mu
- \frac{1}{2} \pi^\mu \pi_\mu
\end{equation}
where $b_\mu$ and $\pi_\mu$ are additional auxiliary fields.
Since they appear without derivatives in the Lagrangian, they can be
eliminated by means of their equations of motion which yield

\begin{equation}
{\it L}_{gf}  = \frac{1}{2} F_\mu F^\mu + \frac{1}{2} f_\mu f^\mu
\end{equation}
where $F_\mu$ is given in equation (\ref{firstghost}) and $f_\mu$ has
the following form \cite{MKL2}:

\begin{equation}
f_\lambda = \left(f_1 g_{\sigma \lambda} g^{\mu \nu}
+  f_2 \delta^\mu_\lambda \delta^\nu_\sigma
+ f_3 \delta^\nu_\lambda \delta^\mu_\sigma \right)
\gamma^\sigma_{~\mu \nu}
\label{general}
\end{equation}
where $\{f_i\}$ are constants satisfying the condition

\begin{equation}
f_1 + f_2 + 4f_3 \neq 0
\end{equation}

For constructing the quantum Lagrangian we must add the appropriate
Faddeev-Popov ghost fields. We derive the corresponding theory from the
invariance of the full Lagrangian under the BRST-transformation

\begin{equation}
{\it s} {\it L}_{quan}  = 0
\end{equation}
where {\it s} is a nilpotent BRST operator.
In the background field formalism we violate the symmetry connected with
the transformation of quantum fields. The BRST-transformation is
obtained in the usual way from gauge transformation by replacing the
gauge parameter by the corresponding ghost field. The complete
BRST-transformations for all fields are the following:

\begin{eqnarray}
{\it s} g_{\mu \nu} & = & 0
{}~~~~~~~~~~~~{\it s} ç^\sigma_{~\mu \nu} =  0 \nonumber \\
{\it s} h_{\mu \nu}  & = &
 \left(\nabla_\mu c_\nu + \nabla_\nu c_\mu \right)
+  {\it k} \left(
c^\lambda \nabla_\lambda h_{\mu \nu} + \nabla_\mu c^\lambda
h_{\lambda \nu} + \nabla_\nu c^\lambda h_{\lambda \mu}
\right) + O ({\it k}^2)
\nonumber \\
{\it s} c^\mu &  = & c^\lambda \partial_\lambda c^\mu
{}~~~~~{\it s} \bar{c}^\mu  =  b^\mu
{}~~~~~{\it s} b^\mu  =  0
\nonumber \\
{\it s} \bar{\chi}^\mu & = & \pi^\mu
{}~~~~~~~~~~{\it s} \pi^\mu = 0
{}~~~~~~{\it s} \chi_\nu = 0
\nonumber \\
{\it s} \gamma^\sigma_{~\mu \nu}(x) & = &
 D^\sigma_{~\alpha \nu} \nabla_\mu c^\alpha
+ D^\sigma_{~\mu \alpha } \nabla_\nu c^\alpha
- D^\alpha_{~\mu \nu} \nabla_\alpha c^\sigma
+ c^\alpha \nabla_\alpha D^\sigma_{~\mu \nu}
\nonumber  \\
& + & \frac{1}{2} \left( \nabla_\mu \nabla_\nu + \nabla_\nu \nabla_\mu
\right) c^\sigma
 - \frac{1}{2} \left( R^\sigma_{~\mu \nu \beta} + R^\sigma_{~\nu \mu
\beta} \right) c^\beta - \delta^\sigma_\mu \chi_\nu
\nonumber \\
& - & {\it k} \left( \gamma^\alpha_{~\mu \nu} \nabla_\alpha c^\sigma
- \gamma^\sigma_{~\alpha \nu} \nabla_\mu c^\alpha
- \gamma^\sigma_{~\mu \alpha } \nabla_\nu c^\alpha
- c^\alpha \nabla_\alpha \gamma^\sigma_{~\mu \nu} \right)
+ O ({\it k}^2)
\label{BRST}
\end{eqnarray}
where $\left( \bar{c}_\mu, c_\nu \right)$ and
$\left( \bar{\chi}_\mu, \chi_\nu \right)$ are the anticommuting ghost fields
connected with general coordinate and projective transformations,
respectively.

The quantum Lagrangian is

\begin{equation}
{\it L}_{quan}  = L_{eff}(\gamma, h) + {\it s}
\biggl\{ \bar{c}^\mu \left(F_\mu - \frac{1}{2} b_\mu \right)
+ \bar{\chi}^\mu \left( f_\mu - \frac{1}{2} \pi_\mu \right)
\biggr\}
\label{full-action}
\end{equation}
where $L_{eff}(\gamma, h)$ is the action (\ref{action}) expanded as a
power series in the quantum fields. This, together with the condition
${\it s}^2 = 0$, implies immediately the BRST invariance of the action
(\ref{full-action}).

 From (\ref{BRST}) and (\ref{full-action}) we obtain the one-loop ghost
Lagrangian

\begin{eqnarray}
{\it L}_{gh}  & = &   - \bar{c}^\mu {\it s} F_\mu
- \bar{\chi}^\mu {\it s} f_\mu  \nonumber \\
& = &
- \left( \bar{c}^\mu ~\bar{\chi}^\mu \right)
\left( \begin{array}{cc}
 \left(g_{\mu \nu} \nabla^2 + R_{\mu \nu} \right) & 0 \\
Z_{\mu \nu} &  (f_1 + f_2 + 4f_3) g_{\mu \nu} (-g)^\alpha
\end{array} \right)
\left( \begin{array}{c}
c^\nu \\
\chi^\nu
\end{array} \right)
\label{full-ghost}
\end{eqnarray}
where
$\alpha $ is a constant and

\begin{eqnarray}
Z_{\lambda \sigma} & = &
- f_1 g_{\lambda \sigma} \nabla ^2  - \frac{1}{2} \left( f_2 + f_3 \right)
\left( \nabla_\lambda \nabla_\sigma + \nabla_\sigma \nabla_\lambda \right)
+ \frac{1}{2} \left( f_2 + f_3 - 2 f_1 \right) R_{\lambda \sigma}
\nonumber \\
& - &   \left( f_1 \nabla_\sigma
D^\mu_{\lambda~~ \mu} + f_2 \nabla_\sigma D_\lambda +
f_3 \nabla_\sigma D^\mu_{~\mu \lambda} \right) -
\left( f_2 D_\sigma + f_3 D^\mu_{~\mu \sigma} \right) \nabla_\lambda
\nonumber \\
& + & f_1  \left( g_{\lambda \sigma} D^{\alpha \beta}_{~~~\beta}
  \nabla_\alpha
  - D_{\lambda \sigma}^{~~~\mu} \nabla_\mu
- D_{\lambda ~ \sigma}^{~ \mu} \nabla_\mu  \right)
\end{eqnarray}

The following relations are valid for arbitrary triangular matrix
operator:
\begin{eqnarray}
\ln {\it det}
\left( \begin{array}{cc}
 A & 0 \\
C & B
\end{array} \right)
& = &
{\it Sp} \ln
\left( \begin{array}{cc}
 A & 0 \\
C & B
\end{array} \right)
 =
{\it Sp}
\left( \begin{array}{cc}
 \ln A & 0 \\
 K & \ln B
\end{array} \right)
\nonumber \\
 & = & \ln A + \ln B =
\ln {\it det}
\left( \begin{array}{cc}
 A & 0 \\
 0 & B
\end{array} \right)
\label{rel}
\end{eqnarray}
where A, B and C are arbitrary operators.

Using these relations we can write the one-loop ghost contribution to
the effective action in a more convenient form

\begin{equation}
ç_{gh}  =
-i \ln {\it det} \left( g_{\mu \nu} \nabla^2 + R_{\mu \nu} \right)
-i \ln {\it det} \left( (f_1 + f_2 + 4f_3) g_{\mu \nu} (-g)^\alpha
\right)
\label{GH1}
\end{equation}
where the first and second terms are the one-loop contribution of the
coordinate and projective ghosts, respectively.

The validity of the relation (\ref{GH1}) can also be proven in a
different way. To get the diagonal form of the ghost Lagrangian
(\ref{full-ghost}) we define a new field

\begin{equation}
\tilde {\chi}_\nu = \chi_\nu + Z_{\nu \sigma} c^\sigma
\end{equation}

This redefinition does not change the functional measure. In the  new
variables the ghost Lagrangian has the diagonal form

\begin{equation}
{\it L}_{gh}
- \left( \bar{c}^\mu ~\bar{\chi}^\mu \right)
\left( \begin{array}{cc}
 \left(g_{\mu \nu} \nabla^2 + R_{\mu \nu} \right) & 0 \\
 0 &  (f_1 + f_2 + 4f_3) g_{\mu \nu} (-g)^\alpha
\end{array} \right)
\left( \begin{array}{c}
c^\nu \\
\tilde {\chi}^\nu
\end{array} \right)
\end{equation}
The ghost contribution to the one-loop effective action is given by the
relation (\ref{GH1}).

To simplify our calculation we use the following projective
gauge condition instead of (\ref{general}):

\begin{equation}
f_\lambda =
A \delta^\beta_\sigma \delta^\alpha_\lambda
\gamma^\sigma_{~\alpha \beta} \equiv
f_{\lambda  \sigma }^{~~~\alpha  \beta }
\gamma^\sigma_{~\alpha  \beta }
\label{prgf}
\end{equation}
where the constant $A$ is nonzero.

The one-loop contribution of the projective ghosts to the effective
action is proportional to  $\delta^4(0)$. In the dimensional
regularization $[\delta^4(0)]_R = 0$, and the contribution of the
projective ghosts to the one-loop counterterms is equal to zero.

Now, we must change equation (\ref{second}).
The propagator of the quantum field $\gamma^\sigma_{~\mu \nu }$
satisfies equation (\ref{first}) and the new condition

\begin{equation}
\bar{F}^{-1 \sigma ~~\lambda}_{~~~~\mu \nu ~\alpha \beta}
\overline F_{\lambda~~\rho}^{~\alpha \beta ~ \tau \epsilon} =
\delta^\sigma_\rho \delta^\tau_\mu \delta^\epsilon_\nu
\label{instead}
\end{equation}

where

\begin{eqnarray}
\overline F_{\sigma ~~\lambda}^{~\alpha \beta ~\mu \nu} & = &
\tilde F_{\sigma ~~\lambda}^{~\alpha \beta ~\mu \nu} +
f_{\tau \sigma }^{~~~\alpha \beta } f^{\tau ~~\mu \nu }_{~\lambda }
\nonumber \\
& = &
      b_1 g_{\sigma \lambda } g^{\beta \nu } g^{\alpha \mu }
  -   b_1 g_{\sigma \lambda } g^{\beta \mu } g^{\alpha \nu }
- \left(1 - \frac{b_2}{2} \right)
g^{\nu \alpha}\delta^\mu_\sigma \delta^\beta_\lambda
- \left(1 - \frac{b_2}{2}\right)
g^{\mu \beta} \delta^\nu_\sigma \delta^\alpha_\lambda
\nonumber \\
& + &
\left(A^2 + \frac{b_3}{2}\right) g^{\alpha \mu} \delta^\nu_\lambda
 \delta^\beta_\sigma  - \frac{b_3}{2} g^{\mu \beta}
 \delta^\alpha_\sigma \delta^\nu_\lambda - \frac{b_3}{2} g^{\alpha \nu}
\delta^\beta_\sigma \delta^\mu_\lambda + \frac{b_3}{2} g^{\nu
\beta}\delta^\alpha_\sigma \delta^\mu_\lambda
\nonumber \\
& + &
g^{\mu \nu} \delta^\alpha_\lambda \delta^\beta_\sigma + g^{\alpha \beta}
\delta^\mu_\sigma \delta^\nu_\lambda - \frac{b_2}{2} g^{\alpha \mu }
\delta^\beta_\lambda \delta^\nu_\sigma - \frac{b_2}{2} g^{\beta \nu }
\delta^\alpha_\lambda \delta^\mu_\sigma
\end{eqnarray}

Having solved equations (\ref{first}) and (\ref{instead}) we obtain the
following result:

\begin{eqnarray}
\overline{F}^{-1 \alpha ~~\mu}_{~~~~\beta \sigma ~\nu\lambda}  =
 & - & \frac{1}{4}  g^{\alpha \mu} g_{\beta \sigma} g_{\nu \lambda} +
  A_2 g^{\alpha \mu} g_{\beta \nu} g_{\sigma \lambda}
+ \biggl( \frac{1}{12} - \frac{1}{3} \left(A_2 - A_1 \right)
\biggr)
g_{\nu \beta} \delta^\mu_\lambda \delta^\alpha_\sigma
\nonumber \\
& + & \frac{1}{4}  \biggl( g_{\nu \lambda} \delta^\mu_\beta
\delta^\alpha_\sigma
+ g_{\beta \sigma} \delta^\alpha_\nu \delta^\mu_\lambda \biggr)
 + A_1 \biggl( g_{\nu \sigma}
\delta^\alpha_\lambda \delta^\mu_\beta + g_{\beta \lambda}
\delta^\mu_\sigma \delta^\alpha_\nu \biggr)
\nonumber \\
& - &
\frac{1}{4}
\biggl( g_{\nu \lambda}
\delta^\mu_\sigma \delta^\alpha_\beta + g_{\beta \sigma}
\delta^\alpha_\lambda \delta^\mu_\nu \biggr)
- \left(A_2 - \frac{1}{2} \right)
g^{\alpha \mu } g_{\sigma \nu } g_{\beta \lambda }
\nonumber \\
& + &
\biggr( \frac{1}{3} \left(A_2 - A_1 \right) - \frac{1}{12} \biggr)
\biggl( g_{\beta \lambda}
\delta^\mu_\nu \delta^\alpha_\sigma + g_{\beta \nu} \delta^\alpha_\beta
\delta^\mu_\lambda \biggr) \nonumber \\
\nonumber \\
& + &
\biggl( \frac{1}{12}  - \frac{1}{3} \left(A_2 - A_1 \right) +
A^2 \biggr)
g_{\sigma \lambda} \delta^\mu_\nu \delta^\alpha_\beta
\nonumber \\
& - &
\left( \frac{1}{2}  + A_1 \right)
g_{\nu \beta } \delta^\mu_\sigma \delta^\alpha_\lambda  -
\left( \frac{1}{2}  + A_1 \right)
g_{\sigma \lambda } \delta^\mu_\beta \delta^\alpha_\nu
\end{eqnarray}
where the constants $A_2, A_1$ are defined from expressions
(\ref{coef}).

The abandonment calculations coincide with the previous
case. Having made the replacement of the variables (\ref{replace}), one
needs to change   $ F^{-1 \alpha ~~\mu}_{~~~~\beta \sigma ~\nu\lambda}
\rightarrow
\overline{F}^{-1 \alpha ~~\mu}_{~~~~\beta \sigma ~\nu\lambda}$.
We fix the coordinate invariance  by the conditions
(\ref{firstghost}) and (\ref{secondghost}), and the Lagrangian of the
coordinate ghosts is defined by  (\ref{coorghost}). The one-loop
counterterms on mass-shell coincide with expression (\ref{result}).

\section{Conclusion}

In the present paper, we have investigated the role of the terms
quadratic in the torsion fields at the quantum level in the theory with
independent metric and connection fields. It turns out that:

\begin{enumerate}
\item  In the affine metric theory the terms quadratic in the torsion
fields play an auxiliary role. They serve for violating projective
invariance of the action.
\item  The renormalizability of the model (\ref{action}) is not affected
by the presence of the projective invariance.
\item  In the considered model (\ref{action}) the terms quadratic in the
torsion fields do not contribute to the one-loop counterterms.
\end{enumerate}

Let us consider the additional conditions (\ref{d}) arising in the
definition of the quantum field propagator. It is easy to show that

\begin{equation}
d = 8(1-b_1-b_2)(1+2b_1 - b_2)
\end{equation}
The coefficients $(1-b_1-b_2)$ and $(1+2b_1 - b_2)$ are proportional to
the particle masses arising in the linear field approximation
{}~\cite{HS,Blag1,Blag2}.  The condition $d = 0$
corresponds to the presence of massless particles in the theory.
In this case, the propagator of the quantum field $\gamma^\sigma_{\mu
\nu }$ is not defined. Hence, the appearance of  new massless
particles is connected with the presence of the new type symmetry in
the theory. We do not known the exact transformation rule of the fields
under these new symmetries ~\cite{Blag2,Bars,Sun}.  It
is known that the connection field is transformed under these
symmetries. The metric field is not changed.

The theory involved is renormalizable at the one-loop level on the
mass-shell. The expression $ \int d^4x \sqrt{-g} \biggl(R_{\alpha \beta
\mu \nu}R^{\alpha \beta \mu \nu} - 4 R_{\mu \nu } R^{\mu  \nu} + R^2
\biggr) $ is proportional  to  the topological number of
space-time, the so-called Euler number, defined as

\begin{equation}
\chi   =  \frac{1}{32 \pi^2 }
 \int d^4x \sqrt{-g} \biggl(R_{\alpha \beta \mu
\nu}R^{\alpha \beta \mu \nu} - 4 R_{\mu \nu } R^{\mu  \nu} + R^2
\biggr)
\end{equation}

Hence, this expression is some number. In the topological trivial
space-time this number is equal to zero. Then, at the one-loop level on
mass-shell one needs to renormalize only the cosmological constant. Let
us represent the cosmological constant in the following form:

\begin{equation}
\Lambda  = \frac{\lambda }{{\it k}^2}
\end{equation}
where $\lambda $  is the dimensionless constant. Then, from the explicit
calculations in the previous section, we get the renormalization group
equation

\begin{equation}
\beta (\bar{\lambda})
= \mu^2 \frac{\partial \bar{\lambda} }{\partial \mu^2 }
= -\frac{29}{160 \pi^2 } \bar{\lambda}^2
\end{equation}
where $\mu^2 $  is the renormalization point mass.  Hence, we have the
asymptotic freedom for $\lambda $.

The result of the one-loop calculations on mass-shell
coincides with the one-loop counterterms of the Einstein gravity
with the cosmological constant~\cite{Chris}. This coincidence is
accidental. The considered theory coincides with the Einstein
gravity at the tree level.  In the nonrenormalizable theories the
results of the loop calculations depend on the choice of the
dynamical variables. It has been shown ~\cite{MKL3} that the
classical theory written in a different way leads to
inequivalent quantum results. Since the Einstein gravity and theory
under consideration are not renormalizable at the two-loop level, the
equivalence of the above-mentioned theories can be violated at the
quantum level.  Therefore, one cannot  predict the result of the
one-loop calculations (\ref{result}) without the corresponding
calculations.

We are greatly indebted to L.O.Vasilyeva, A.Gladyshev and G.
Sandukovskaya for  critical reading of the manuscript.

\end{document}